\documentclass[twocolumn]{aastex63}
\usepackage{amsmath}
\usepackage{lineno}

\newcommand{\g}{\gamma}

\newcommand{\psim}{\lower.5ex\hbox{$\; \buildrel \propto \over\sim \;$}}
\newcommand{\lbar}{\lower.0ex\hbox{$\; \buildrel{\lower0.0ex \hbox{-}} \over\lambda  \;$}}

\newcommand{\Epk}{E_{\rm pk}}

\newcommand{\cm}{\mathrm{cm}}
\newcommand{\km}{\mathrm{km}}
\newcommand{\erg}{\mathrm{erg}}

\newcommand{\keV}{\mathrm{keV}}
\newcommand{\MeV}{\mathrm{MeV}}

\newcommand{\s}{\mathrm{s}}

\newcommand{\Mpc}{\mathrm{Mpc}}

\shorttitle{SIRI-2 Detection of the GRB 221009A}
\shortauthors{Mitchell et al.}

\begin{document}

\title{SIRI-2 Detection of the Gamma-Ray Burst 221009A}
\author{Lee J.\ Mitchell}
\email{lee.j.mitchell5.civ@us.navy.mil}
\author{Justin D.\ Finke}
\email{justin.d.finke.civ@us.navy.mil}
\author{Bernard Phlips}
\affil{U.S.\ Naval Research Laboratory, Code 7650, 4555 Overlook Ave.\ SW,
        Washington, DC,
        20375-5352, USA 
        }

\author{W.\ Neil Johnson}
\author{Emily Kong}
\affil{Technology Service Corporation, 251 18th Street South, Suite 705, Arlington, VA, 22202, USA}

\begin{abstract}

SIRI-2 is a collection of Strontium Iodide gamma-ray detectors sensitive at approximately 400 keV to 10 MeV, launched on the Department of Defense's STPSat-6 to geosynchronous orbit. SIRI-2 detected the gamma-ray burst (GRB) 221009A and, unlike most GRB detectors, was not saturated and did not require any pulse pile-up corrections.  The energetics of this burst as measured by SIRI-2 are consistent with those found by other instruments, and the Band function fits to the spectra are consistent with that from the unsaturated Insight and GECAM instruments, and similar to corrected spectra from the {\em Fermi} Gamma-ray Burst Monitor, but softer than those found by Konus-Wind when that instrument was saturated.  The total fluence measured with SIRI-2 was measured to be $0.140\pm0.002\ \erg\ \cm^{-2}$, lower than other instruments, likely due to the increasing background of SIRI-2 forcing the calculation to use a smaller time interval.  An extrapolation of the distributions of fluences from GRBs to the fluence of 221009A measured with SIRI-2 indicates bursts brighter than this one should occur about once every 4,000 years.

\end{abstract}


\section{Introduction}
\label{intro}

The brightest gamma-ray burst (GRB) ever detected, 221009A (also known as Swift J1913.1+1946), was triggered on 2022 October 9 at 13:16:59.99 UTC by {\em Fermi} Gamma-ray Burst Monitor \citep[GBM;][]{veres22,lesage23}.  It was also detected by other well-established instruments, including {\em Fermi} Large Area Telescope \citep{bissaldi22,pillera22,axelsson24}, Swift Burst Alert Telescope \citep[BAT;][]{kennea22,williams23}, Konus-Wind \citep{frederiks23}, {\em AGILE} \citep{tavani23}, MAXI \citep{negoro22}, and INTEGRAL \citep{rodi23}.  It was seen unsaturated in the 10 keV to 6 MeV range by Insight-HXMT and GEACAM-C instrument \citep{an23}.  It was the first ever GRB to be detected by a water Cherenkov detector, namely by LHAASO \citep{cao23,cao23_2}.  Its redshift was found to be $z=0.151$ \citep{deugarte22,izzo22,malesani23}.  Many authors have used this to explore the possibility of or constraints on new physics, e.g., Lorentz invariance violation \citep[e.g.,][]{finke23,vardanyan23,piran23,lhaaso23,li24} or axion-like particles \citep[e.g.,][]{carenza22,baktash22,troitsky22,nakagawa23,wang23,zhang23,galanti23,bernal23,troitsky24}.  The detection of $\g$-rays from this burst scattered by the atmosphere allowed for the constraint of the $\g$-ray polarization of the burst \citep{veres24_polar}.  No neutrinos were detected by IceCube coincident with this burst \citep{veres24_neutrino}.

This GRB was dubbed the ``Brightest Of All Time'' or ``BOAT'' \citep{burns23}, since it has the highest peak flux, fluence, and isotropic equivalent energy release of any GRB; and one of the highest peak luminosities.  However, these brightness estimates were made based on estimated corrections for detector saturation, since GRB 221009A saturated almost every $\g$-ray detector which observed it.  One exception is the Strontium Iodide Radiation Instrument II \citep[SIRI-2;][]{mitchell22}.  This lack of saturation allows SIRI-2 to probe this GRB's spectrum during the brightest parts of its $\g$-ray light curve, at a time when other instruments could not observe it without substantial corrections for pulse pile-up.  This allows for a view of GRB 221009A that does not require complicated and possibly error-prone pulse pile-up corrections. It allows for a cross-check of these analysis and detection techniques, and for SIRI-2 observations to possibly validate the corrections.

Here we describe the SIRI-2 observations of GRB 221009A.  In Section \ref{SIRI_section} we give a short description of the SIRI-2 instrument itself.  In Section \ref{observations_section} we describe the SIRI-2 observations of this GRB, including spectral fitting.  Finally we conclude with a discussion in Section \ref{discussion_section}.

\section{SIRI-2 Instrument}
\label{SIRI_section}

So far, two SIRI instruments have been launched as part of the Department of Defense (DoD) Space Test Program \citep[e.g.,][]{sims09}.  The original SIRI instrument was launched on 2018 Dec 3 to low Earth orbit on board STPSat-5 satellite \citep{mitchell17}.  It consisted of a single europium-doped strontium iodide (SrI$_2$:Eu)  crystal in an aluminum enclosure with a total mass of 1.62 kg.  The scintillating light was read out with a silicon photomultiplier (SiPM).  The SiPM provides similar performance to traditional photo-multipliers, but with a lower bias, power, and space.  The  SrI$_2$:Eu provides greater spectral resolution than most room temperature scintillators currently used in $\g$-ray astronomy, and without the cryogen requirements of, e.g., Germanium used by COSI \citep{tomsick23}.  COSI, of course, will have other benefits, such as the ability to track particles reconstruct event directions and measure polarization.

SIRI-2 was launched onboard STPSat-6 on 2021 Dec 7 into a geosynchronous orbit (GEO) and has been in operation since 2022 Jan 14.  SIRI-2’s primary mission objective is to demonstrate the performance of SrI$_2$:Eu gamma-ray detection technology with sufficient active area for DoD operational needs. The original mission duration was one year and has been extended to three years, with the potential for seven. The instrument’s science objective is to study the gamma-ray emission of solar flares in the energy range of 500-7000 keV and is loosely based on the design of the Solar Maximum Mission which operated throughout the 1980s \citep{singer65}.  SIRI-2 was designed to spectral features in solar flares, such as the nuclear de-excitation lines, generated when accelerated ions become excited after interacting with ambient solar materials. The de-excitation process of an individual nucleus typically occurs in less than $10^{-10}$\ s and ranges in energy between 1--8 MeV. Solar flares can also generate a significant low-energy Bremsstrahlung radiation component during an event, which can lead to saturation of the instrument and reduce the instruments capability to measure the higher energy de-excitation lines. Using a variety of passive and active shields, SIRI-2 was designed specifically to measure the MeV de-excitation lines without saturating on the low energy component of the solar flare; the primary detector allows SIRI-2 to observe these high energy photons during rare intense solar flares. Similarly, the bulk of photons in a GRBs are also low-energy in nature ($<300$\ keV) and in the case of GRB221009A the shear brightness of the event has saturated the large volume, high efficiency instruments (in the MeV range) with low energy thresholds. 

\begin{figure}
\vspace{2.2mm} 
\epsscale{1.1} 
\plotone{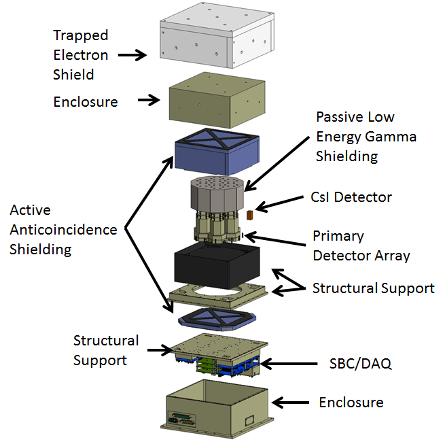}
\caption{An exploded view showing the major components of the SIRI-2 instrument.}
\label{SIRI2exploded_fig}
\vspace{2.2mm}
\end{figure}

Figure \ref{SIRI2exploded_fig} shows an exploded view of SIRI-2 instrument. The main components include a trapped electron shield, an anti-coincidence  detector (ACD), a passive low energy shield, a data acquisition system and the detectors. SIRI-2’s primary detector array has a frontal area of 66 cm$^2$ and consists of 7 hexagonal shaped, SrI$_2$:Eu detectors. The energy resolution is ~4\% at 662 keV for each detector and they are arranged in a close-packed arrangement.  Individual detector crystal dimensions are 3.81 cm in diameter and 3.81 cm tall. 

Onboard multi-channel analyzers (MCAs)  can provide event histogramming, event-by-event list mode data, and individual pulse shape measurements or pulse ``scope traces''. Individual detector rates are logged every two seconds and the value represents the number of events that occurred in that detector over that time interval. Spectral histograms (1024 bins) are continuously logged every 10 seconds. The gain is set to bin events up to ~7 MeV; events greater in energy are recorded by the overflow bin. The low-energy threshold is set to $\sim400$\ keV. In the intense radiation background of GEO, this insures the data acquisition system is not saturated or the allotted bandwidth is not exceeded.  The current configuration is optimized for solar flare transients, which tend to be substantially larger in duration than GRBs. Further details of SIRI-2 can be found elsewhere \citep{mitchell19}.

\section{SIRI-2 Observations of GRB 221009A}
\label{observations_section}

\subsection{Detection of GRB 221009A}

\begin{figure*}
\vspace{2.2mm} 
\epsscale{1.1} 
\plotone{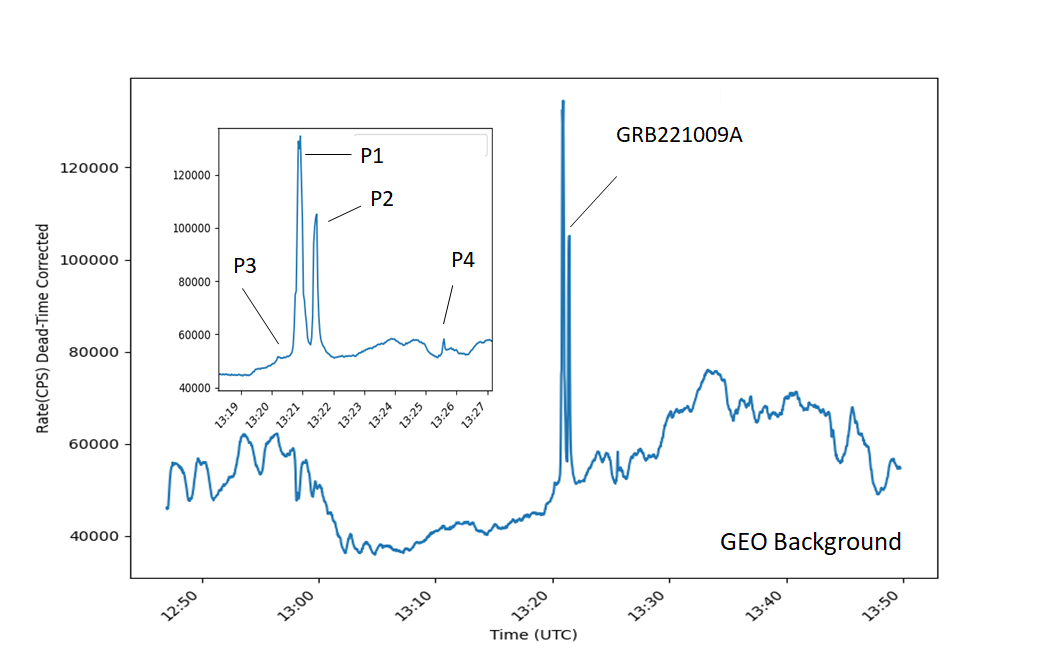}
\caption{The sum of event rates in the 7 SIRI-2 Strontium Iodide detectors on 2022 October 9. Four peaks were observed by SIRI-2: the primary peak (P1) and secondary peaks ( P2-P4).}
\label{221009ALC_fig}
\vspace{2.2mm}
\end{figure*}

SIRI-2 detected the bright long-duration GRB 221009A; this instrument does not have trigger alert algorithm, so it was only noticed once the GRB had been announced by other instruments.  Figure \ref{221009ALC_fig} shows the light curve as measured by the SIRI-2 instrument and was generated from the state-of-health messages which log the count rate in each detector every two seconds. The figure also shows the intense and highly variable background of GEO. Bremsstrahlung radiation is generated by trapped electrons interacting with the spacecraft.  SIRI-2 has difficulty detecting typical GRBs due to this background; however, GRB 221009A was bright enough to still be observable.  From the light curve, four peaks are easily identifiable and have also been observed by {\em Fermi}-GBM \citep{lesage23}. The largest two peaks P1 and P2 in the light curve had peak event rates of $\approx90,000$ and $\approx70,000$ count s$^{-1}$ during that 2 second interval. These values have been corrected for dead time which peaked at $\sim50$\% during the time of P1. Smaller peaks in the light curve, P3 and P4 where also observed in SIRI-2 instrument, and correspond temporally to event peaks seen in the  {\em Fermi}-GBM data \citep{lesage23}. Note that the naming convention here is different than the one used by the GBM Collaboration \citep{lesage23}.

\begin{figure*}
\vspace{2.2mm} 
\epsscale{1.1} 
\plotone{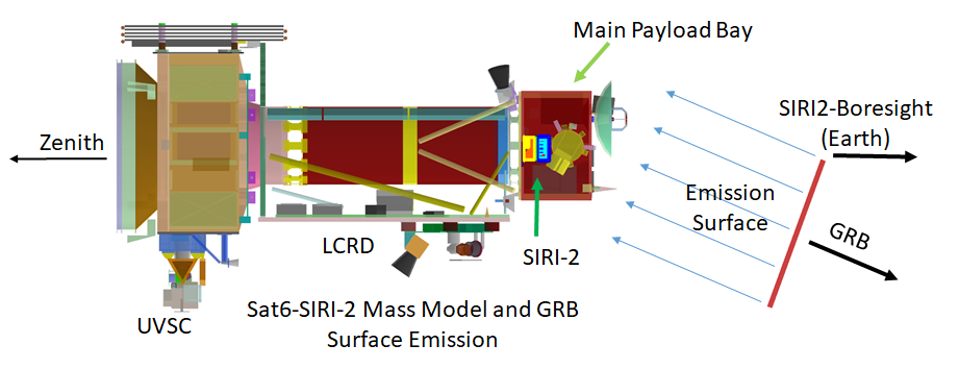}
\caption{The mass model of the STPSat-6 used in the SWORD-Geant4 simulations of GRB 221009A.  The direction of the GRB and the Earth are shown in the illustration.}
\label{SIRI2sword_fig}
\vspace{2.2mm}
\end{figure*}

The direction of the GRB relative to the SIRI-2 instrument was determined based upon coordinates found by the Jansky Very Large Array \citep{laskar22}.  Using spacecraft ephemeris data it was determined that the GRB was $19.97\arcdeg$ off-boresight of the instrument. This was extremely fortunate for two reasons:  the SIRI-2 instrument is pointed towards the Earth; and the FOV is occulted by the payload electronics and the spacecraft on the sides and aft  (see Figure \ref{SIRI2sword_fig}). From the 10-second histograms, the background-subtracted count rate between 400-7000 keV averaged 65,000 count s$^{-1}$ and 24,000 count s$^{-1}$ from 2022 Oct 9 13:20:51-13:21:11 and 13:21:11-13:21:41 UTC, respectively. This coincided temporally with the previously-reported peaks in the GRB light curve \citep[e.g.,][]{lesage23,frederiks23}.

\subsection{Simulations}

There are no response functions for SIRI-2.  Instead we simulated the GRB using a SoftWare for Optimization of Radiation Detection \citep[SWORD;][]{duvall19} model for SIRI-2.  SWORD is a graphical user interface that allows easy access to the Geant4 \citep{agostinelli03} and Monte Carlo N-Particle \citep{werner18} radiation transport codes.  
Figure \ref{SIRI2sword_fig} shows the mass model of the STPSat6 spacecraft used in the simulations.
GRB spectra are usually described by a Band function \citep{band93}
\begin{equation}
\label{bandfcn}
\frac{dN}{dE}  = A \begin{cases}
E_2^\alpha e^{-E_2/E_{s,2}} & E_2\le(\alpha-\beta)E_{s,2} \\
E_2^\beta (\alpha-\beta)E_{s,2}^{\alpha-\beta}e^{\beta-\alpha} &  (\alpha-\beta)E_{s,2} < E_2
\end{cases}
\ ,
\end{equation}
where $E_2$ is the observed photon energy in units $10^2\ \keV$.  This
has free parameters $A$, $\alpha$, $\beta$, and
$\Epk\equiv(2+\alpha)E_{s,2}\times10^2\ \keV$.  With SWORD and Geant4 we simulated the GRB with several values of the Band function parameters $\alpha$, $\beta$\, and $E_{\rm pk}$.  With these models, we can interpolate between the simulated instrument spectrum to get the simulated instrument spectrum for any set of parameter values.

\subsection{Spectra}

\begin{figure}
\vspace{2.2mm} 
\epsscale{1.1} 
\plotone{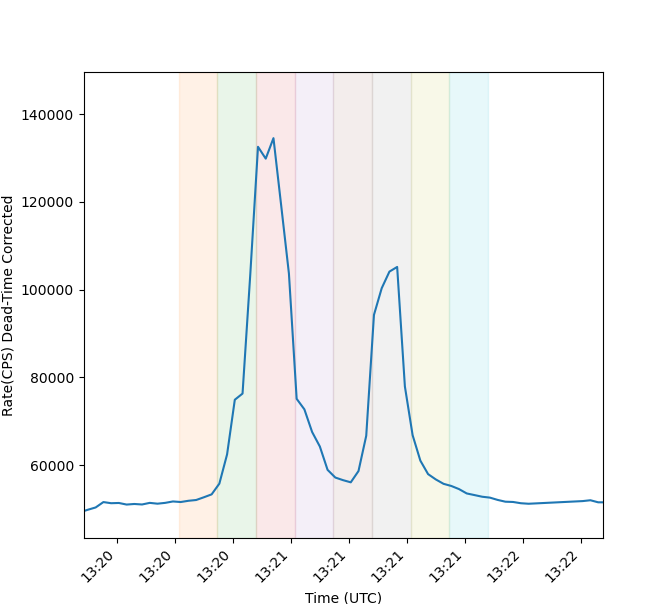}
\caption{The event rate light curve for GRB 221009A as observed by SIRI-2. 
 The sequential histograms automatically generated by the SIRI-3 analysis and used for the spectral fits are shown as the shaded regions of various colors.}
\label{lightcurve2_fig}
\vspace{2.2mm}
\end{figure}

Using the interpolation between the simulated instrument spectra, we performed a Markov chain Monte Carlo fit to the SIRI spectrum for GRB 221009A in 10 second intervals.  This gives us the parameter values and 68\% confidence intervals for the model parameters.  For comparison, we measure all times relative to the {\em Fermi}-GBM trigger from this GRB \citep[2022 October 9 at 13:16:59.99 UTC;][]{lesage23}.  The model parameter results from our fits can be found in Table \ref{spectratable}.  The time intervals are shown as the shaded regions on the light curve in Figure \ref{lightcurve2_fig}.  Unfortunately, the time intervals are decided before the analysis was done; the user does not have the ability to choose the time interval sizes and locations.  In cases where we got an upper limit on $E_{\rm pk}$ due to the limited energy range of the spectrum, we fixed the value of $\alpha$.  Although fits were performed in the 600 keV -- 7 MeV range, we extrapolated the spectrum outside this range to get the 20 keV -- 10 MeV flux.  

\begin{deluxetable*}{lcccc}
\tablecaption{Result of Band Function fits
}
\tablewidth{0pt}
\tablehead{
\colhead{Time since GBM trigger [s]} &
\colhead{$\alpha$} &
\colhead{$\beta$} &
\colhead{$E_{\rm pk}$ [keV]} &
\colhead{20 keV - 10 MeV flux [$\erg\ \cm^{-2}\ \s^{-1}$]}
}
\startdata
$211-221$ & $-1.1$\tablenotemark{*} & $-2.08\pm0.02$ & $<860$ & $(1.6\pm0.6)\times10^{-5}$ \\
$221-231$ & $-1.28\pm0.03$ & $-2.06\pm0.01$ & $831\pm46$ & $(7.2\pm0.2)\times10^{-3}$ \\
$231-241$ & $-0.91\pm0.03$ & $-2.47\pm0.01$ & $788\pm22$ & $(4.9\pm0.1)\times10^{-3}$ \\
$241-251$ & $-1.1$\tablenotemark{*} & $-2.69\pm0.01$ & $<724$ & $(4.60\pm0.07)\times10^{-5}$ \\
$251-261$ & $-1.1$\tablenotemark{*} & $-2.35\pm0.01$ & $<763$ & $(7.3\pm0.1)\times10^{-4}$ \\
$261-271$ & $-0.90\pm0.014$ & $-2.45\pm0.02$ & $746\pm29$ & $(1.36\pm0.04)\times10^{-3}$ \\
$271-281$ & $-1.1$\tablenotemark{*} & $-2.30\pm0.10$ & $<775$ & $(1.43\pm0.03)\times10^{-5}$ \\
$281-291$ & $-1.04\pm0.06$ & $-2.00\pm0.10$ & $1951\pm160$ & $(1.64\pm0.05)\times10^{-6}$ \\
\enddata
\tablenotetext{*}{Fixed in fit.}
\label{spectratable}
\end{deluxetable*}

\begin{figure}
\vspace{2.2mm} 
\epsscale{1.25} 
\plotone{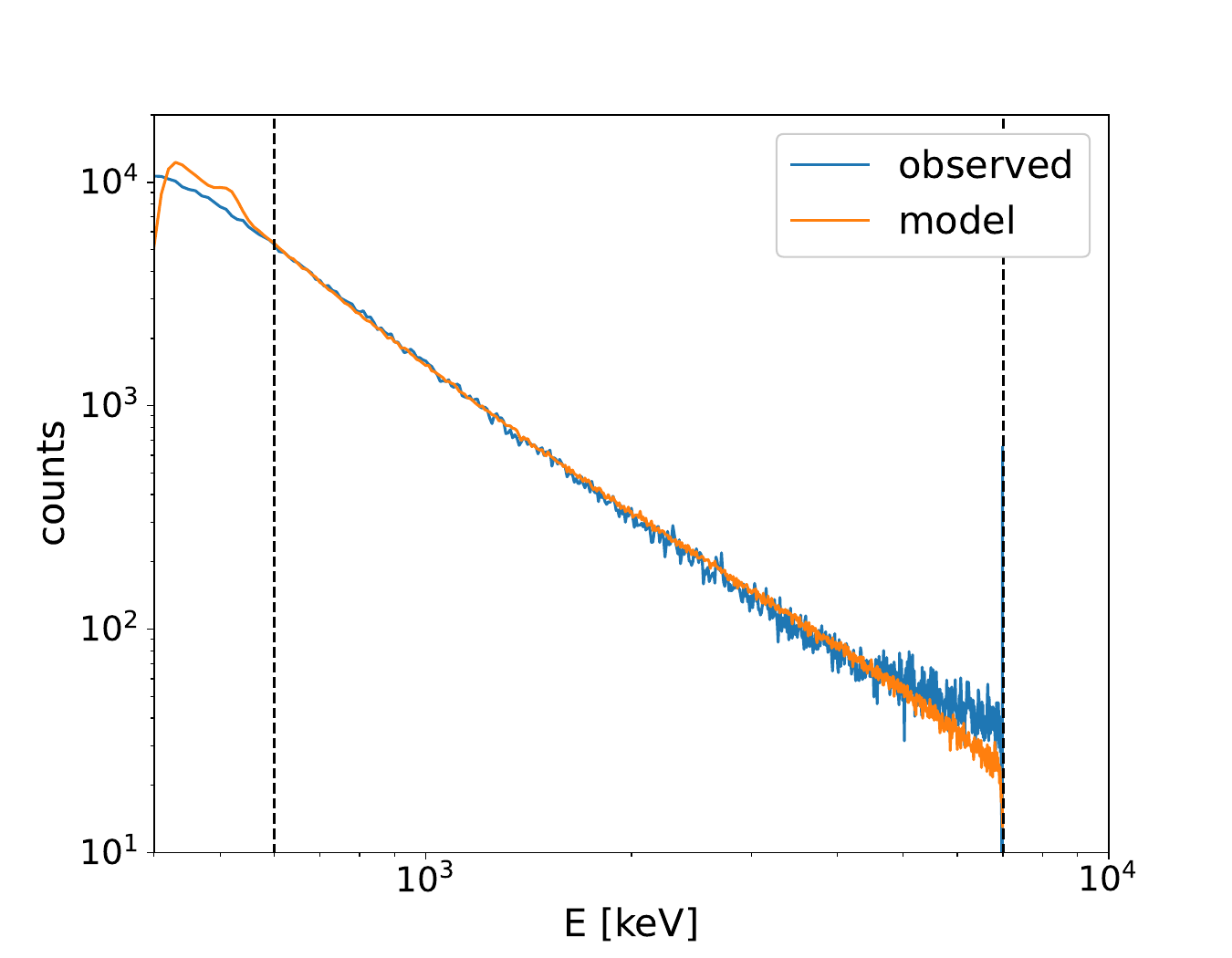}
\caption{SIRI-2 count spectrum for 221009A for the time interval 221-231 s after the GBM trigger.  The observed spectrum and the best fit model are shown.  The dashed lines indicate the energy range used in the fit.}
\label{spectrumfig}
\vspace{2.2mm}
\end{figure}

An example of a count spectrum and its best fit model are shown in Figure \ref{spectrumfig}.  The fit was only performed in the 600 keV to 7 MeV range.  Below 600 keV the the simulations show photoelectrons not present in the physical spectra.  Above 5 MeV, the observed spectrum shows a small ``upturn'', which is not well modeled by the simulation.  There is some uncertainty in the mass model of the STPSat spacecraft, since some of its payload is classified and unavailable to us.  This is likely the cause of the model not perfectly describing the observed spectra.

\begin{figure*}
\vspace{2.2mm} 
\epsscale{1.1} 
\plottwo{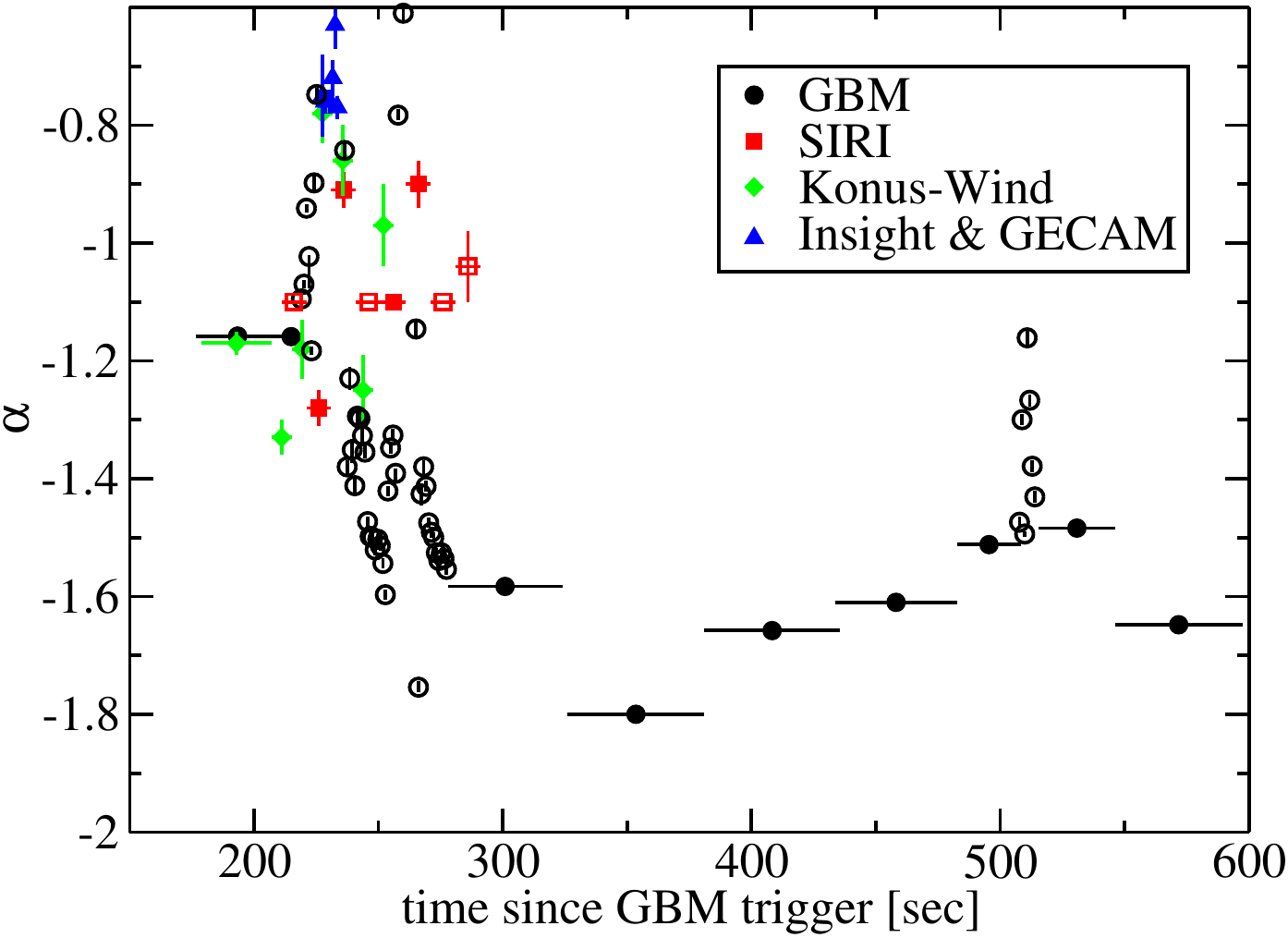}{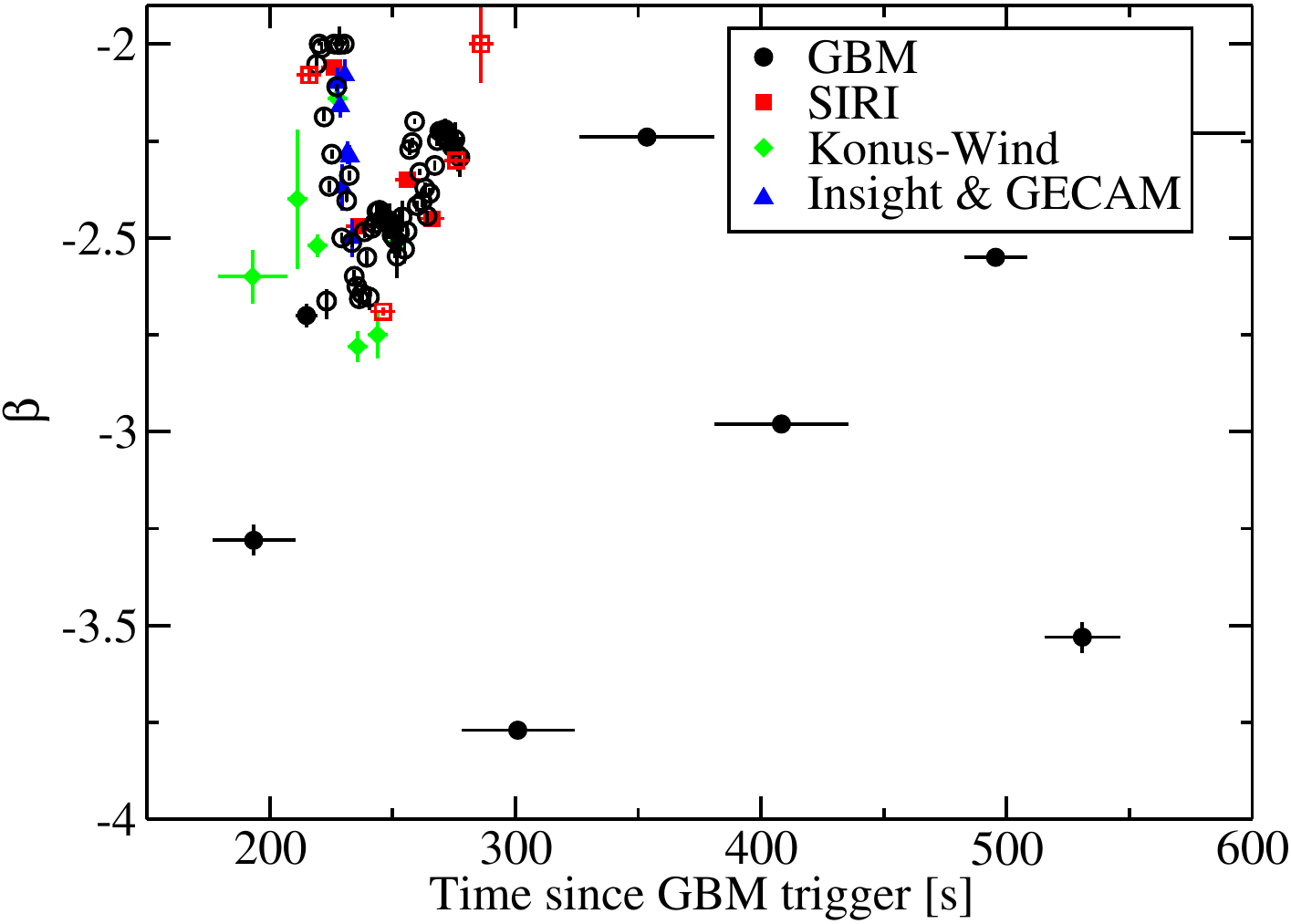}
\caption{The Band parameters $\alpha$ ({\em left}) and $\beta$ ({\em right}) as a function of time for GRB 221009A.  The different instruments are shown with the legend.  Filled black circles are the GBM data from \citet{lesage23}, while the empty black circles are the GBM data from \citet{veres24_neutrino}.  Filled red squares indicate high-precision results for SIRI-2, while empty red squares indicate low-precision results.  }
\label{bandalphabeta}
\vspace{2.2mm}
\end{figure*}

\begin{figure*}
\vspace{2.2mm} 
\epsscale{1.1} 
\plottwo{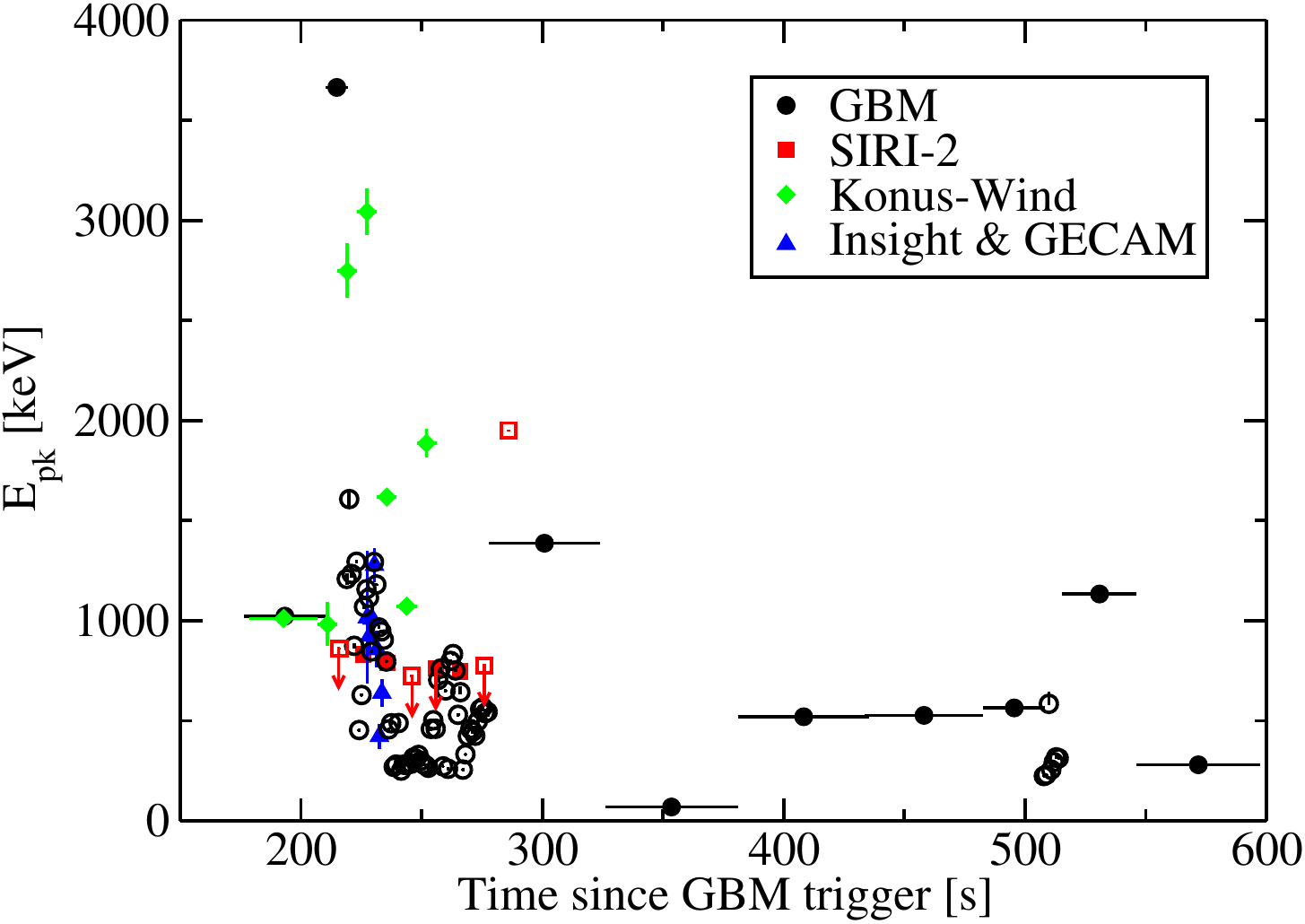}{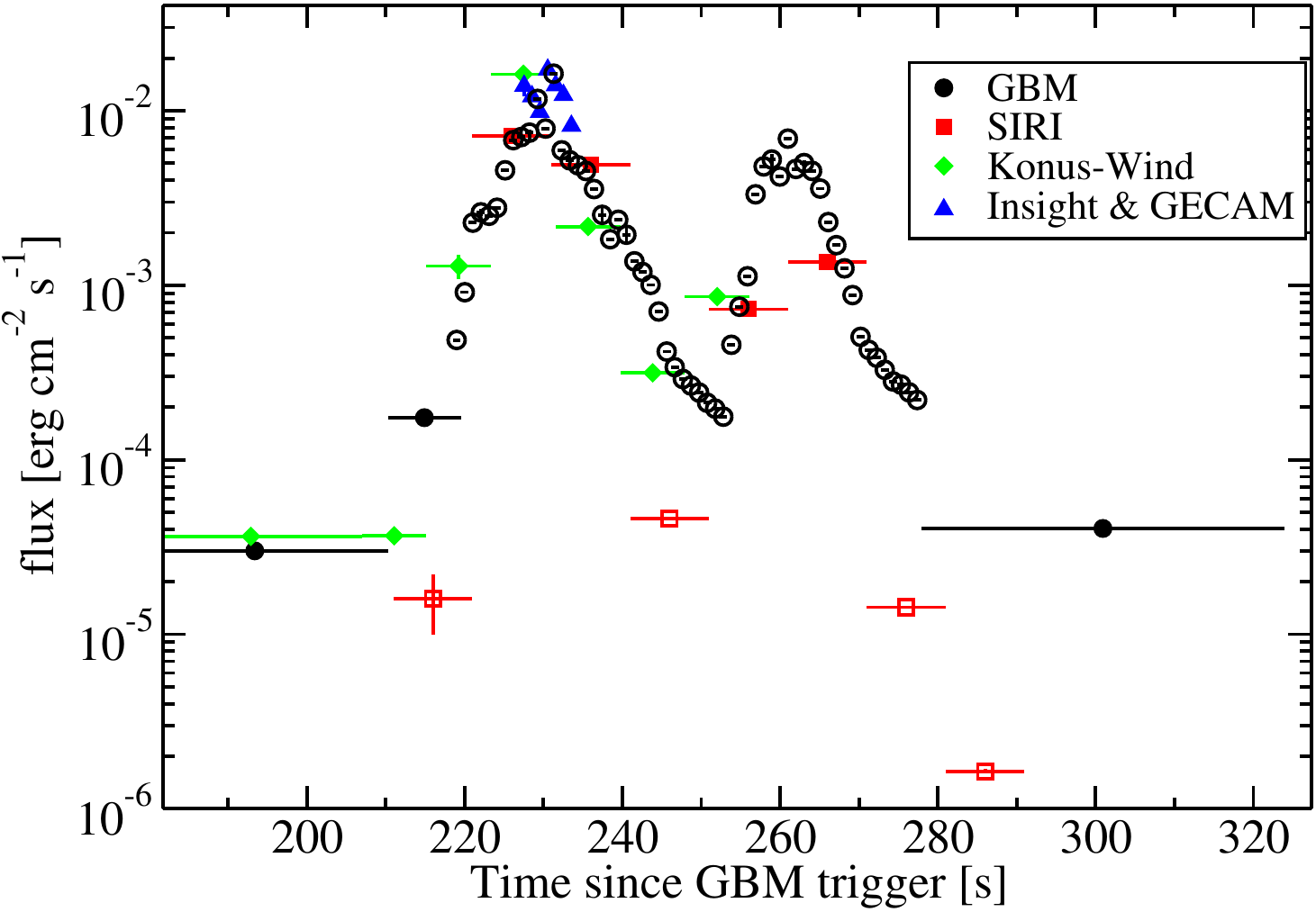}
\caption{The Band parameter $E_{\rm pk}$ ({\em left}) and flux ({\em right}) from the Band function fit as a function of time for GRB 221009A.  Symbols are the same as in Figure \ref{bandalphabeta}.  }
\label{bandEpkflux}
\vspace{2.2mm}
\end{figure*}

Spectral results for SIRI-2, along with results from {\em Fermi}-GBM \citep{lesage23,veres24_neutrino}, Konus-Wind \citep{frederiks23}, and Insight and GECAM \citep{an23} are plotted in Figure \ref{bandalphabeta} and \ref{bandEpkflux}.   \citet{veres24_neutrino} provide spectral results during the ``bad time intervals'' when GBM was saturated, using pulse pile-up corrected spectra.  \citet{frederiks23} also estimate spectral parameters for Wind with spectra corrected for pulse pile-up.  For GECAM and SIRI-2, no corrections for pulse pile-up were needed.  SIRI-2 results at low flux are less reliable, due to the difficulty in subtracting the large, highly variable background (see Figure \ref{221009ALC_fig}); these results are symbolized by the empty red squares.  The values of $\alpha$ do not, in general, have good agreement between the instruments; however, the agreement with values of $\beta$ are much better.  The poor agreement in $\alpha$ may be due to the smaller energy range at low energies.   The values of $E_{\rm pk}$ obtained with SIRI-2, GBM, and Insight/GECAM agree reasonably well, although these are much lower than the values obtained by Konus-Wind  in the 200-250 s time interval after the GBM trigger.  This may be because there were some issues with the pulse pile-up corrections in the time region performed by \citet{frederiks23}.  The flux values obtained by all of these instruments agree quite well, except for the low flux value results for SIRI-2.  As discussed above, difficulty in subtracting the large and highly variable background at GEO makes these results less reliable.  

Basic synchrotron theory prescribes that if the gamma-ray emission is from synchrotron radiation, the lower energy index $\alpha \le -2/3$, the so-called ``line-of-death'' \citep{preece98}.  All of the values of $\alpha$ measured by SIRI-2 and other instruments for GRB 221009A (Figure \ref{bandalphabeta}) are consistent with this constraint.

\section{Discussion}
\label{discussion_section}

We have described the SIRI-2 instrument, and its observation of the GRB 221009A.  Unlike other $\g$-ray detectors such as {\em Fermi}-GBM and Konus-Wind that observed the burst, SIRI-2 did not experience any pulse pile-up effects during the burst, and no corrections were needed. 
Overall energetics observed by SIRI-2 are consistent with those observed by other instruments. The spectral parameters are rather different than those observed by Konus-Wind.  To the extend the results for the different instruments agree, the SIRI-2 observations generally validate of the pulse pile-up corrections for Konus-Wind and {\em Fermi}-GBM during the saturated periods.  

From these results, we can compute the implied 1 keV -- 10 MeV fluence by extrapolating the Band function outside of the SIRI-2 bandpass, giving $0.140\pm0.002\ \erg\ \cm^{-2}$.  This is a bit less than the values found by Konus-Wind, $0.22\ \erg\ \cm^{-2}$; and by GBM\footnote{The manuscript by \citet{lesage23} incorrectly states the fluence measured by GBM is $0.0947\pm0.007\ \erg\ \cm^{-2}$.  This is a mistake based on an earlier analysis (S.\ Lesage, private communication).}, $0.19\ \erg\ \cm^{-2}$.  Our fluence values only include the two peaks labeled P1 and P2 in Figure \ref{221009ALC_fig}, i.e., the time between 221 s and 291 s after the GBM trigger.  The background is too high outside of this region to make a reliable determination on the amount of emission from the burst.  Our fluence value can thus be considered a lower limit, and is compatible with fluence values measured by other detectors.

\citet{ravasio23} claim to have  found a line feature in the GBM spectrum of 221009A at $\ga 10\ \MeV$ \citep[see also][]{axelsson24}.  The energy range of SIRI-2 does not extend beyond 7 MeV, so it is unable to independently verify this measurement.

Using this fluence measured with SIRI-2, a redshift of $z=0.15$, and a flat $\Lambda$CDM cosmology with $H_0=69.6\ \km\ \s^{-1}\ \Mpc^{-1}$ and $\Omega_m=0.286$ \citep{bennett14}, we compute the $E_{\rm iso}$ for 221009A to be $(7.4\pm 0.1)\times10^{54}\ \erg$.  A correlation was found between $E_{\rm pk,i}=(1+z)E_{\rm pk}$ and $E_{\rm iso}$ by \citet{amati02}.  With an updated data set, \citet{tsvet21} found the relation to be 
$$
\log_{10}\left(\frac{E_{\rm pk,i}}{\rm keV}\right) = a\log_{10}\left(\frac{E_{\rm iso}}{\rm erg}\right) + b\ ,
$$
with $a=(0.429\pm0.002)$ and $b=(-20.18\pm0.12)$.  Using our value for $E_{\rm iso}$, and the value of $E_{\rm pk}$ from the 221-231 s interval when the burst is brightest (Table \ref{spectratable}), we find our result for 221009A is $\approx1.9\sigma$ away from this correlation.  If we used a brighter value for $E_{\rm iso}$, assuming some of the flux is missed by SIRI-2, we would get a value even farther from the Amati relation.  This is contrary to the result with Konus-Wind \citep{frederiks23}, where a higher $E_{\rm pk}$ was found; with this higher $E_{\rm pk}$, 221009A was completely consistent with the Amati relation as found by \citet{tsvet21}.

\begin{figure}
\vspace{2.2mm} 
\epsscale{1.25} 
\plotone{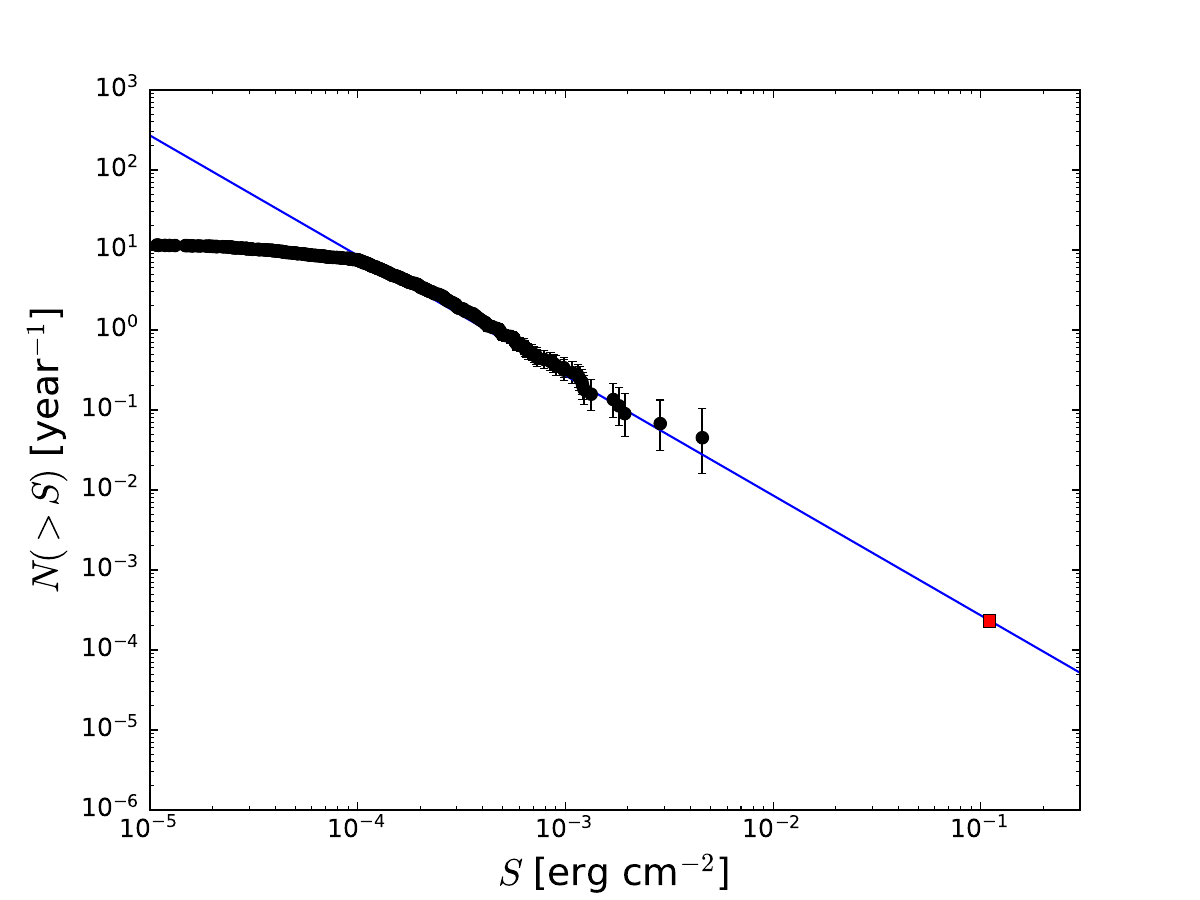}
\caption{The ``logN-logS'' plot for long GRBs, i.e., the number of long GRBs above the given 1 keV -- 10 MeV fluence.  The circles are the total GRBs from all instruments, from \citet{burns23}, with 68\% Poisson uncertainties.  The blue line is $N(>S)\propto S^{-3/2}$ that is normalized to match the data.  The red square indicates the fluence of 221009A, as measured by SIRI-2.}
\label{logNlogSplot}
\vspace{2.2mm}
\end{figure}

In Figure \ref{logNlogSplot} we plot the distribution of long GRBs for all instruments, as found by \citet{burns23}.  We have included 68\% Poisson uncertainties, computed following \citet{gehrels86}.  Extrapolating this population to the SIRI-2 fluence for 221009A, we find that GRBs as bright or brighter than it should occur about once every 4,000 years.  \citet{burns23} used this plot and the fluence of 221009A from Konus-Wind to determine that bursts as bright or brighter than 221009A should have a rate of approximately one per 10,000 years.  However, if there is another population of nearby ($z\la 0.4$) GRBs with narrow jet opening angles, such GRBs could occur at a rate as high as one per 200 years \citep{finke24}.  Alternatively, the extreme brightness of 221009A could be caused by gravitational lensing \citep{bloom24}.  Understanding the energetics of 221009A is crucial to understanding the rate of such extreme GRBs.  SIRI-2 provides an alternative, unsaturated view of this fascinating burst.

\acknowledgments
We are grateful to the referee for useful comments that have improved this paper; to S.\ Lesage and P. Veres for discussions about GRB 221009A with {\em Fermi}-GBM,  E.\ Burns for the logN-logS data plotted here, and S.\ Razzaque for general discussions about GRB 221009A.  This work was partially supported by the Office of Naval Research.  The DoD Space Test Program provided support for SIRI-2 spacecraft, launch, instrument integration, and operations.

\bibliographystyle{apj}
\bibliography{grb_ref,references,gravwave_ref,mypapers_ref,liv_ref,blazar_ref}

\end{document}